\documentclass[english,aps,pra,superscriptaddress,twocolumn,showpacs]{revtex4}
\usepackage[T1]{fontenc}
\usepackage[latin9]{inputenc}
\usepackage{color}
\usepackage{amsmath}
\usepackage{amssymb}
\usepackage{graphicx}
\usepackage{bm,bbm}

\makeatletter
\@ifundefined{textcolor}{}
{%
 \definecolor{BLACK}{gray}{0}
 \definecolor{WHITE}{gray}{1}
 \definecolor{RED}{rgb}{1,0,0}
 \definecolor{GREEN}{rgb}{0,1,0}
 \definecolor{BLUE}{rgb}{0,0,1}
 \definecolor{CYAN}{cmyk}{1,0,0,0}
 \definecolor{MAGENTA}{cmyk}{0,1,0,0}
 \definecolor{YELLOW}{cmyk}{0,0,1,0}
}

\newcommand{\scalar}[2]{\langle #1 | #2 \rangle}

\newcommand{\ket}[1]{| #1 \rangle}
\newcommand{\bra}[1]{\langle #1 |}

\newtheorem{theorem}{Theorem}
\newtheorem{lemma}{Lemma}

\makeatother

\usepackage{babel}
\begin{document}

\title{Strong Majorization Entropic Uncertainty Relations}

\author{\L{}ukasz Rudnicki}

\email{rudnicki@cft.edu.pl}

\affiliation{Freiburg Institute for Advanced Studies, Albert-Ludwigs University
of Freiburg, Albertstrasse 19, 79104 Freiburg, Germany}

\affiliation{Center for Theoretical Physics, Polish Academy of Sciences, Aleja
Lotnik{\'o}w 32/46, PL-02-668 Warsaw, Poland}

\author{Zbigniew Pucha\l{}a}

\affiliation{Institute of Theoretical and Applied Informatics, Polish Academy
of Sciences, Ba\l{}tycka 5, 44-100 Gliwice, Poland}

\affiliation{Institute of Physics, Jagiellonian University, ul Reymonta 4, 30-059
Krak{\'o}w, Poland}

\author{Karol \.{Z}yczkowski}

\affiliation{Institute of Physics, Jagiellonian University, ul Reymonta 4, 30-059
Krak{\'o}w, Poland}

\affiliation{Center for Theoretical Physics, Polish Academy of Sciences, Aleja
Lotnik{\'o}w 32/46, PL-02-668 Warsaw, Poland}

\date{April 30, 2014}

\begin{abstract}
We analyze  entropic uncertainty relations in a finite
dimensional Hilbert space and derive several strong bounds
for the sum of two entropies obtained in projective measurements
with respect to any two orthogonal bases. 
We improve the recent bounds by Coles and Piani, 
which are known to be stronger than the well known result of Maassen and Uffink.
Furthermore, we find a novel bound based on majorization techniques,
which also happens to be stronger than the recent results involving largest singular values 
of submatrices of the unitary matrix connecting both bases.
The first set of new bounds give better results for unitary matrices close
to the Fourier matrix, while the second one provides a significant improvement in the opposite sectors.
Some results derived admit generalization to arbitrary mixed states, so that
corresponding bounds are increased by the von Neumann entropy of the measured state. The majorization approach is finally extended to the case of several measurements.
\end{abstract}
\maketitle

\section{Introduction}


Fundamental differences between the classical and quantum physics
are highlighted by quantum uncertainty relations.
Original version of the relations by Heisenberg, Kennard and Robertson
deal with the sum of uncertainties characterizing two measurements
of observables which do not commute. Right hand sides of these inequalities
are proportional to the size of the Planck constant $\hbar$
as in the classical case the bounds tend to zero.

In the following paper we focus on probably the most popular representatives
of uncertainty relations there are nowadays, given in terms of information
entropies. One uses the standard Shannon entropy, with a clear operational 
meaning, or generalized quantities of  R{\'e}nyi and Tsallis (for reviews on entropic uncertainty relations see \cite{Wehner,IBBLR}). 
One may observe a growing interest in these issues 
of the community working in the theory of quantum information processing \cite{Ra12,my,oni,Coles,ZBP13}
and in applications to for example quantum memory \cite{Berta} or
Einstein\textendash{}Podolsky\textendash{}Rosen steering inequalities
\cite{EPR}. Our aim is thus to classify recent improvements of various
entropic uncertainty relations and provide several new results outperforming
the previous ones.

Before we start let us introduce the notation. For a probability distribution
$p=\left\{ p_{i}\right\} $ its R{\'e}nyi entropy of order $\alpha$ is
given by the formula
\begin{equation}
H_{\alpha}\left(p\right)=\frac{1}{1-\alpha}\ln\sum_{i}p_{i}^{\alpha}.
\end{equation}
In the limit $\alpha\rightarrow1$ the above definition recovers the
Shannon entropy $H\left(p\right)=-\sum_{i}p_{i}\ln p_{i}$. Looking
from a general perspective, R{\'e}nyi entropies of any order are Schur-concave
functions. In fact, every function $F\left(p\right)$ which is Schur-concave
is in position to be a reasonable measure of uncertainty since it
is maximized by a uniform probability distribution, while its minimum
is provided by concentrated probabilities $p_{c}^{\downarrow}=\left(1,0,\ldots,0\right)$.
The symbol $\downarrow$ denotes the decreasing order, so that $\left(p^{\downarrow}\right)_{i}\geq\left(p^{\downarrow}\right)_{j}$
whenever $i\leq j$. Among other Schur-concave functions let us only
mention the so-called  Havrda--Charvat--Tsallis entropy \cite{HC67}
\begin{equation}\label{eqn:tsallis}
T_{\alpha}\left(p\right)=\frac{1}{1-\alpha}\left(\sum_{i}p_{i}^{\alpha}-1\right).
\end{equation}

Describing the quantum state of the system we shall use a mixed state
$\rho$ acting on a $d$-dimensional Hilbert space $\mathcal{H}$.
We will consider two non-degenerate, non-commuting observables $\hat{A}$
and $\hat{B}$ with corresponding eigenstates denoted by $\left|a_{i}\right\rangle $
and $\left|b_{j}\right\rangle $ respectively. The above eigenstates
obviously provide two orthonormal bases in $\mathcal{H}$. We then
define the probability distributions in a usual manner:
\begin{equation}\label{eqn:def-p-q}
p_{i}=\left\langle a_{i}\right|\rho\left|a_{i}\right\rangle ,\qquad q_{j}=\left\langle b_{j}\right|\rho\left|b_{j}\right\rangle .
\end{equation}

The history of the entropic uncertainty relations in finite dimensional
Hilbert spaces (continuous case had been developed before \cite{BBM})
started with the paper by Deutsch \cite{Deutsch} who proved that
\begin{equation}
H\left(p\right)+H\left(q\right)\geq-2\ln C\equiv B_{\textrm{D}},\label{De}
\end{equation}
with $C=\left(1+\sqrt{c_{1}}\right)/2$ and $c_{1}=\max_{i,j}\left|\left\langle a_{i}\left|b_{j}\right\rangle \right.\!\right|^{2}$
being the maximal overlap between the bases $\left\{\left|a_{i}\right\rangle\right\}$
and $\left\{\left|b_{j}\right\rangle\right\}$. This seminal but rather weak lower
bound for the sum of two Shannon entropies was further significantly
improved and generalized by Maassen and Uffink in 1988 when they derived
their famous uncertainty relation \cite{MU} 
\begin{equation}
H_{\mu}\left(p\right)+H_{\nu}\left(q\right)\geq-\ln c_{1}\equiv B_{\textrm{MU}},\label{Mu}
\end{equation}
valid however only for conjugated parameters $1/\mu+1/\nu=2$. 
In the case of a single qubit more general bounds
for an arbitrary pair $\mu,\nu$ were recently studied in \cite{ZBP13a}.

A natural range for the parameter $c_{1}$ is of the form $1/d\leq c_{1}\leq1$.
Comparing the both bounds (\ref{De}) and (\ref{Mu}) in two opposite
regimes of $c_{1}$ one can observe that: 
\begin{itemize}
\item The Maassen-Uffink bound is substantially stronger ($B_{\textrm{MU}}\gg B_{\textrm{D}}$)
in the regime of small $c_{1}\gtrsim1/d$, when both bases are almost
mutually unbiased.
\item In the second case when $c_{1}\lesssim1$ both bounds provide almost
the same quantitative description of uncertainty, 
however, the bound  (\ref{Mu}) is
 always a bit stronger than (\ref{De}),  $B_{\textrm{MU}}\gtrsim B_{\textrm{D}}$.
\end{itemize}
In fact, when $c_{1}=1/d$ and the bases  $\left\{\left|a_{i}\right\rangle\right\}$
and $\left\{\left|b_{j}\right\rangle\right\}$ are related via discrete Fourier
transformation, the bound in (\ref{Mu}) equal to $\ln d$ is optimal.
At this place let us mention that the two bases in question are in
general related by a unitary transformation $U\in\mathcal{U}\left(d\right)$,
with matrix elements equal to $U_{ij}=\left\langle a_{i}\left|b_{j}\right\rangle \right.\!$,
so that $c_{1}=\max_{i,j}\left|U_{ij}\right|^{2}$. 

In the \emph{25 years} (1988-2013) mid time only one example of a general
state-independent improvement of the lower bound (\ref{Mu}), valid
and significant in the regime of large $c_{1}$, has been communicated
\cite{deVicente,deVicenteComm}. Several results were however devoted
to particular studies of eg. qubits \cite{Sanchez,GMR03,BPP}, described by the case $d=2$.

This work is organized as follows. After we review the recent progress in the context of entropic uncertainty relations, we derive in Section II the three new, \textit{hybrid} bounds for the sum of two Shannon entropies. To this end we use both techniques of relative-entropy monotonicity \cite{Coles} and the majorization entropic uncertainty relations \cite{my,oni,Partovi}. In Section III we introduce yet another majorization-based approach which happens to outperform the previous one \cite{my,oni} (see Appendix B). After we compare in Section IV all bounds for the sum of two Shannon entropies which are currently available, we extend in Section V the majorization uncertainty relations derived in Section IV to the case of several measurements.

\subsection{Recent results}

Surprisingly, the Maassen-Uffink bound has been recently improved
in the whole range of the parameter $c_{1}$. First of all, Coles
and Piani \cite{Coles} have provided a state independent bound (note
that we use the natural logarithm instead of $\log_{2}$)
\begin{equation}
H\left(p\right)+H\left(q\right)\geq-\ln c_{1}+\left(1-C\right)\ln\frac{c_{1}}{c_{2}}\equiv B_{\textrm{CP1}},\label{ColPia}
\end{equation}
with  the same as before $C=\left(1+\sqrt{c_{1}}\right)/2$ and 
 $c_{2}$ being the second largest value among $\left|U_{ij}\right|^{2}$.
Since $c_{2}\leq c_{1}$ the second term in (\ref{ColPia}) is a non-negative
correction to (\ref{Mu}). The above example shows that the improvements
of (\ref{Mu}) shall rely on more overlaps between the bases. An intermediate
step in the derivation of (\ref{ColPia}) leads to a stronger but
implicit bound \cite{Coles}: 
\begin{equation}
H\left(p\right)+H\left(q\right)\geq\max_{0\leq\kappa\leq1}\lambda_{\min}\left(-2\Delta\right)\equiv B_{\textrm{CP2}}\geq B_{\textrm{CP1}},\label{ColPia-1}
\end{equation}
involving the $d\times d$ matrix $\Delta$ with matrix elements  $\Delta_{mn}$ being
\begin{equation}
\kappa\delta_{mn}\ln\max_{k}\left|U_{mk}\right|+\left(1-\kappa\right)\sum_{j}U_{mj}U_{nj}^{*}\ln\max_{k}\left|U_{kj}\right|.
\end{equation}
$\lambda_{\min}\left(\cdot\right)$ denotes here the minimum eigenvalue. The above results have been derived only in the case of the Shannon entropy,
since they utilize the relative entropy:
\begin{equation}
D\left(\rho\left|\right|\sigma\right)=\textrm{Tr}\rho\ln\rho-\textrm{Tr}\sigma\ln\sigma.
\end{equation}

While both bounds (\ref{ColPia}, \ref{ColPia-1}) are never worse
than $B_{\textrm{MU}}$, they seem to provide more accurate improvements
for $c_{1}\gtrsim1/d$ {[}note the factor $1-C$ in (\ref{ColPia}){]}
rather than in the case $c_{1}\lesssim1$. In this second regime
another approach based on majorization techniques comes into play.
The idea that majorization can be used to quantify uncertainty \cite{Partovi}
has been developed in \cite{my,oni} giving a bound 
\begin{equation}
F\left(p\otimes q\right)\geq F\left(Q\right),
\end{equation}
valid for any Schur-concave function $F$. By $Q$ we denote any vector
of probabilities that majorizes $r\prec Q$ the distribution $r=p\otimes q$
(we shall call the above result the \emph{tensor-product majorization
relation}). 
Unless otherwise stated we assume that the vector $Q$ is sorted in a decreasing order. The majorization relation $r\prec Q$ means that for all $n<d{}^{2}$
we necessarily have $\sum_{k=1}^{n}r^{\downarrow}_{k}\leq\sum_{k=1}^{n}Q_{k}$
and due to the probability conservation $\sum_{k=1}^{d^{2}}r_{k}=\sum_{k=1}^{d^{2}}Q_{k}=1$.
As long as $c_{1}<1$ there exist nontrivial vectors $Q\neq Q^{\left(0\right)}=\left(1,0,\ldots,0\right)$.
It also happens that the majorizing probability vector $Q$ possesses
at most $d$ nonzero elements. In \cite{my} we derived a full
hierarchy of $d-1$ majorizing vectors $Q^{(k)}$, $k=1,\ldots,d-1$,
such that 
\begin{equation}
Q^{\left(0\right)}\succ Q^{\left(1\right)}\succ Q^{\left(2\right)}\succ\ldots\succ Q^{\left(d-1\right)}\succ r,\label{hier}
\end{equation}
which are expressed by singular values of certain submatrices selected
from the $d\times d$ unitary matrix $U$. 

In particular, an additivity property of the R{\'e}nyi entropies
$H_{\alpha}\left(p\otimes q\right)=H_{\alpha}\left(p\right)+H_{\alpha}\left(q\right)$
immediately provide the bound utilizing the majorizing vector presented in~\cite{my}:
\begin{equation}
H_{\alpha}\left(p\right)+H_{\alpha}\left(q\right)\geq H_{\alpha}\left(Q^{\left(d-1\right)}\right)\equiv B_{\textrm{Maj1}}.
\label{maj1}
\end{equation}
While in the regime $c_{1}\gtrsim1/d$
the bound $B_{\textrm{Maj1}}$ might be weaker than (\ref{Mu}), it
happens that for $d=5$, it is stronger than the result of Maassen
and Uffink with a probability larger than $98\%$ \cite{my}.

\section{Strong bounds for the Shannon entropy}

In the following section we derive three state--independent bounds $B_{\textrm{RPZ}k}(\rho)$, $k=1,2,3$
for the sum of two Shannon entropies utilizing both the relative entropy
approach and the majorization technique. We shall call the bounds \textit{state independent}, they however all contain a non-negative state dependent term equal to the von-Neumann entropy $S(\rho)\geq0$. To get state independent bounds in a strict meaning  (denoted consistently by $B_{\textrm{RPZ}k}(\phi)$) one shall simply chose the state $\rho$ to be pure.

We start recalling first
steps from the derivation of (\ref{ColPia-1}) used by Coles and Piani \cite{Coles}
which concern an arbitrary initial state $\rho$,
\begin{align}
H\left(q\right)+& \textrm{Tr}\rho\ln\rho =  D\left(\rho\left\Vert \sum_{j}q_{j}\left|b_{j}\right\rangle \left\langle b_{j}\right|\right.\right)\label{relmon}\\
 \geq\; & D\left(\sum_{i}p_{i}\left|a_{i}\right\rangle \left\langle a_{i}\right|\left\Vert 
\sum_{j,k}q_{j}\left|U_{jk}\right|^{2}\left|a_{k}\right\rangle \left\langle a_{k}\right|\right.\right).\nonumber 
\end{align}
The inequality is a consequence of monotonicity of the relative-entropy 
with respect to the quantum channel: 
\begin{equation}
\rho\mapsto\sum_{i}\left|a_{i}\right\rangle \left\langle a_{i}\right|\rho\left|a_{i}\right\rangle \left\langle a_{i}\right|.
\end{equation}

We shall now rewrite the above inequality to the form 
\begin{equation}
H\left(p\right)+H\left(q\right)\geq-\sum_{i}p_{i}\ln\left(\sum_{j}q_{j}\left|U_{ij}\right|^{2}\right)+ S\left(\rho\right). \label{part1}
\end{equation}
The second term appearing on the right hand side is equal to the von Neumann entropy $S\left(\rho\right)=-\textrm{Tr}\rho\ln\rho$ of the state $\rho$.  Performing the same step as in (\ref{relmon}), but starting from
$H\left(p\right)$ one derives the following counterpart of (\ref{part1})
\cite{Coles}:
\begin{equation}
H\left(p\right)+H\left(q\right)\geq-\sum_{j}q_{j}\ln\left(\sum_{i}p_{i}\left|U_{ij}\right|^{2}\right)+ S\left(\rho\right).\label{part2}
\end{equation}
\subsection{First application of the tensor-product majorization relation} 

With the help of the convexity property of $-\ln\left(\cdot\right)$
both intermediate bounds (\ref{part1}, \ref{part2}) can be estimated
in the same way, giving 
\begin{equation}\label{inter}
H\left(p\right)+H\left(q\right)\geq-\ln\left(\sum_{i,j}p_{i}q_{j}\left|U_{ij}\right|^{2}\right)+  S\left(\rho\right).
\end{equation}
Let us now denote by $c=\left(c_{1},c_{2},\ldots,c_{d^{2}}\right)$
the $d^{2}$-dimensional vector of elements $\left|U_{ij}\right|^{2}$
sorted in the decreasing order. Recalling the notion of the vector $Q\succ r$ majorizing
the $d^{2}$-dimensional vector $r=p\otimes q$ we immediately get
\begin{equation}
H\left(p\right)+H\left(q\right)\geq-\ln\left(Q\cdot c\right)+ S\left(\rho\right)\equiv B_{\textrm{RPZ1}},
\label{bound1}
\end{equation}
where $Q$ is by definition sorted in the decreasing order
while  $Q\cdot c$ denotes the scalar product of the vectors $Q$ and $c$.
 In order to prove the above result we shall simply notice
that the argument inside the logarithm in (\ref{inter}) is less
than $r\cdot c$ and $-\ln\left(r\cdot c\right)$ is a Schur-concave
function with respect to $r$.

\subsubsection{The simplest estimations} 

Obviously, every $r=p\otimes q$ is majorized by $Q^{\left(0\right)}$.
For that choice the bound (\ref{bound1}) boils down to $B_{\textrm{MU}}$
with a non-negative correction provided by the von Neumann entropy
term $-\textrm{Tr}\rho\ln\rho$.

As a first non-trivial case we can take \cite{my} $Q^{\left(1\right)}=\left(C^{2},1-C^{2},0,\ldots,0\right)$.
This choice leads to a simple and strong, new state independent bound
\begin{equation}
\label{bound2}
H\left(p\right)+H\left(q\right)\geq-\ln\left[c_{1}C^{2}+c_{2}\left(1-C^{2}\right)\right]+S\left(\rho\right)\equiv B_{\textrm{RPZ2}}.
\end{equation}

\subsection{Implicit bounds from the tensor-product majorization relation} 

As a first step we shall take the arithmetic mean of (\ref{part1}) and (\ref{part2}),
and reexpress the resulting inequality in the form
\begin{equation}
H\left(p\right)+H\left(q\right)\geq-\frac{1}{2}\sum_{k,l} p_{k}q_{l}\ln\left(\sum_{i,j}p_{i}q_{j}\left|U_{kj}\right|^{2}\left|U_{il}\right|^{2}\right)+ S\left(\rho\right).
\end{equation}
For each pair of indices $\left(k,l\right)$ we next introduce a
$d^{2}$-dimensional vector $h_{kl}$ given by the elements $\left|U_{kj}\right|^{2}\left|U_{il}\right|^{2}$
sorted in the decreasing order with respect to the pair $\left(i,j\right)$.
We immediately get 

\begin{equation}
-\ln\left(\sum_{i,j}p_{i}q_{j}\left|U_{kj}\right|^{2}\left|U_{il}\right|^{2}\right)\geq-\ln\left(Q\cdot h_{kl}\right).
\end{equation}
In the second step, we introduce the vector $f$ of length $d^2$, given by the
elements $f_{kl} = -\ln\left(Q\cdot h_{kl}\right)$ sorted in the  decreasing
order  with respect to the pair $\left(k,l\right)$. Finally (using the same arguments as before) we obtain the
implicit (two sortings required) relation 
\begin{equation}
\label{bound3}
H\left(p\right)+H\left(q\right)\geq-\frac{1}{2}Q\cdot f+  S\left(\rho\right)\equiv B_{\textrm{RPZ3}}.
\end{equation}

The above considerations enable us to formulate the following hybrid entropic uncertainty relations,
based both on majorization techniques and monotonicity of the relative entropy.
\begin{theorem}
The entropic uncertainity relations presented in (\ref{bound1}), (\ref{bound2}) and (\ref{bound3}),
valid for an arbitrary mixed state $\rho$,  have the form
\begin{equation}
\label{bounds_mixed}
H\left(p\right)+H\left(q\right)\geq B_{\textrm{RPZ}k} (\rho) \ \ge \   B_{\textrm{RPZ}k} (\phi) + S(\rho),
\end{equation}
for $k=1,2,3$.
\end{theorem}

\section{Direct-sum majorization relations}
\label{sec_maj2}

Let $U$ be a unitary matrix of size $d$. 
By $\mathcal{SUB}(U,k)$ we denote the set of all its submatrices of class $k$
defined by 
\begin{eqnarray}
\mathcal{SUB}(U,k)&=&\{M:\#\textrm{cols}(M)+\#\textrm{rows}(M)=k+1\nonumber\\
&&\text{ and }M\text{ is a submatrix of }U\}.
\end{eqnarray}
The symbols $\#\textrm{cols}\left(\cdot\right)$ and $\#\textrm{rows}\left(\cdot\right)$
denote the number of columns and the number of rows respectively.
Following \cite{my} we define the coefficients 
\begin{equation}
s_{k}=\max\left[\|M\|:\; M\in\mathcal{SUB}(U,k)\right],
\label{sk}
\end{equation}
with $\|M\|$ being the operator norm equal to the maximal singular value of $M$.
With this notation we can state the following result.
\begin{theorem}[Direct-sum majorization relation] \label{th:D-S-M}
For the R{\'e}nyi entropies of order $\alpha\leq1$ (and thus also for the Shannon entropy) we have
\begin{equation}
H_{\alpha}(p)+H_{\alpha}(q)\geq H_{\alpha}(W)\equiv B_{\textrm{Maj2}},
\label{Dir}
\end{equation}
where $W=\left(s_{1},s_{2}-s_{1},\dots s_{d}-s_{d-1}\right).$
\end{theorem}

While
the tensor-product majorization \cite{my} is based on the relation $p\otimes q\prec Q$,
its counterpart, giving the meaning to the vector $W$ is of the direct-sum form 
\begin{equation}
\label{WW}
p\oplus q\prec\{1\}\oplus W,
\end{equation}
(see Appendix A for details and the proof of Theorem~\ref{th:D-S-M}). 

In the case $\alpha>1$ the relation (\ref{Dir}) does not hold, we
can however establish a bit weaker bound
\begin{equation}
H_{\alpha}(p)+H_{\alpha}(q)\geq\frac{2}{1-\alpha}\ln\left(\frac{1+\sum_{i}W_{i}^{\alpha}}{2}\right).\label{Renyi2}
\end{equation}
Surprisingly, a relation of the same kind as (\ref{Dir}) holds for
the Tsallis entropy proven in Appendix A.
\begin{theorem}\label{th:Tsallis}
For the Tsallis entropy~\eqref{eqn:tsallis} of \emph{any} order $\alpha\geq0$, we have 
\begin{equation}
T_{\alpha}(p)+T_{\alpha}(q)\geq T_{\alpha}(W).
\end{equation}
\end{theorem}

To give a particular example of the direct-sum majorization entropic
uncertainty relation let us recall that \cite{my} the singular value $s_2$ is upper--bounded in the following way
\begin{equation}
s_{2}\leq\sqrt{c_{1}+c_{2}}.
\end{equation}
We can then use this inequality to provide a particular vector 
\begin{equation}
W^{\left(2\right)}=\left(\sqrt{c_{1}},\sqrt{c_{1}+c_{2}}-\sqrt{c_{1}},1-\sqrt{c_{1}+c_{2}},0,\ldots,0\right),
\end{equation}
that majorizes the vector $W$. This vector contains $2\left(d-2\right)$ zero elements. The above example shows that the same sort of hierarchy as given by
(\ref{hier}) can be constructed in the case of the direct-sum majorization. To this end we simply need to set $W^{\left(1\right)}=\left(\sqrt{c_{1}},1-\sqrt{c_{1}},0,\ldots,0\right)$ and $W^{\left(d-1\right)}=W$. 

In Appendix B we prove yet another majorization relation $W\prec Q^{\left(d-1\right)}$ which
happens to be valid for any unitary matrix $U$. This observation implies
that the bound $B_{\textrm{Maj2}}$ is \emph{always} stronger $B_{\textrm{Maj2}}\geq B_{\textrm{Maj1}}$
than the bounds previously derived in~\cite{my,oni}.

\section{Comparison of bounds}
In this section we illustrate our results showing how the obtained bounds work for selected families of unitary matrices belonging to  $\mathcal{U}(3)$ and $\mathcal{U}(4)$.
We consider first a family of $3 \times 3$ matrices defined as
\begin{equation}\label{eqn:family1}
U(\theta) = M(\theta) O_3 M(\theta)^{\dagger},
\end{equation}
where
\begin{equation}
M(\theta) = 
\left(
\begin{smallmatrix}
1 & 0 & 0 \\
0 & \cos \theta  & \sin \theta \\
0 & - \sin \theta & \cos \theta
\end{smallmatrix}
\right) \  \ \text{ and } \ \
O_3 = 
\frac{1}{\sqrt{6}}
\left(
\begin{smallmatrix}
%
\sqrt{2} & \sqrt{2} & \sqrt{2} \\
\sqrt{3} & 0       & -\sqrt{3} \\
  1      &      -2 &    1
\end{smallmatrix}
\right).
\end{equation}
The matrix $O_3$, used in \cite{Coles} to illustrate the quality of
the bounds $ B_{\textrm{CP1}}$ and $ B_{\textrm{CP2}}$ corresponds to the choice $\theta = 0$.

\begin{figure}
\includegraphics[scale=0.22]{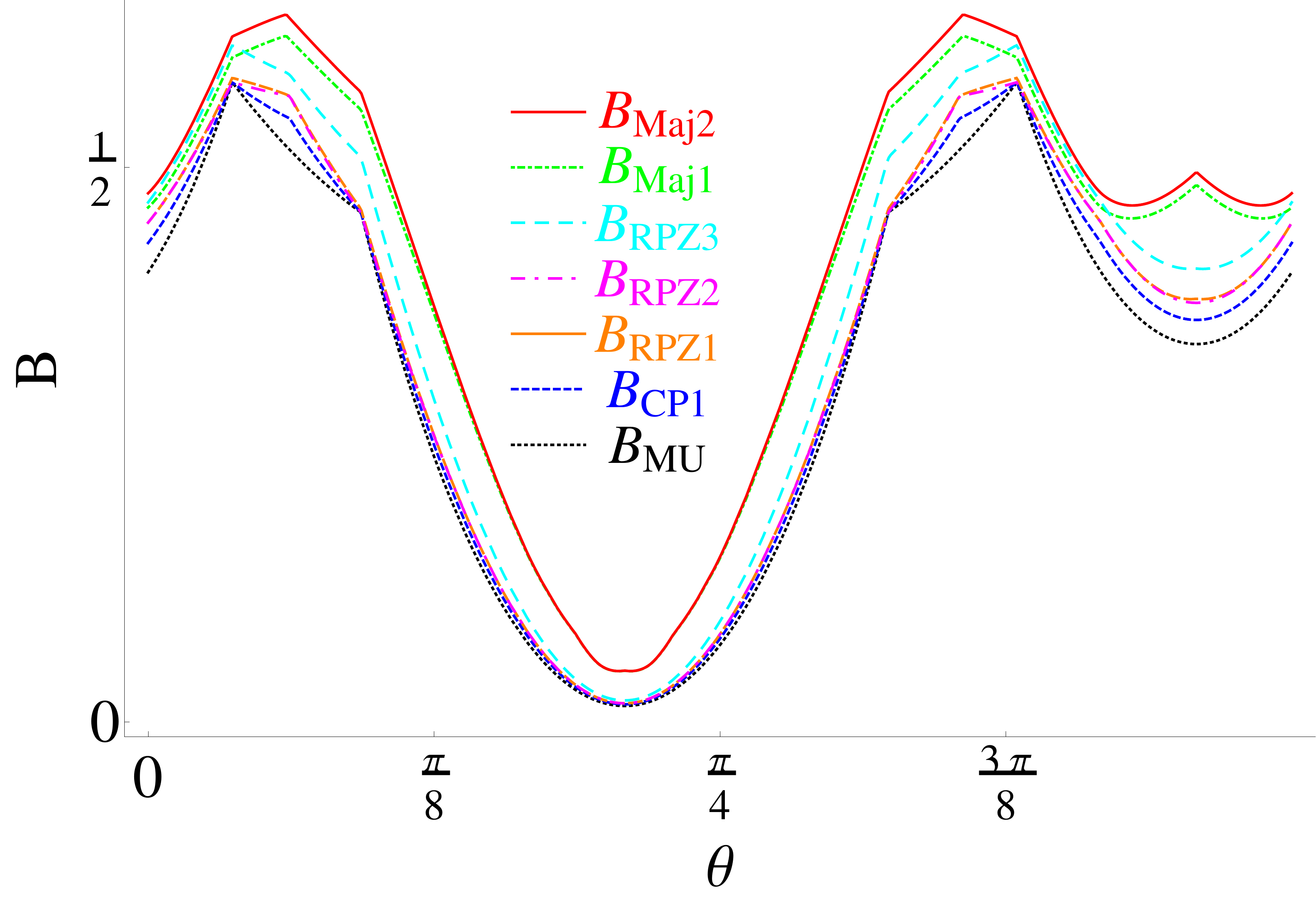} 
\caption{(Color online) Several bounds for the sum of two entropies for a family of unitary matrices defined in~\eqref{eqn:family1}.}
\label{fig:fig1}
\end{figure}

In Fig. \ref{fig:fig1} the bounds discussed in this paper (apart from  $B_{\textrm{CP2}}$ which requires additional numerical optimization) are presented for the example (\ref{eqn:family1}).  The bound $B_{\textrm{Maj2}}$ provides the best estimation for the sum of two
entropies while the majorization bound $B_{\textrm{Maj1}}$ gives (as expected) always a worse
approximation. $B_{\textrm{Maj2}}$ outperforms the Maassen--Uffink bound
as well as the bound $B_{\textrm{CP2}}$. The quantities $B_{\textrm{RPZ1}}$ and $B_{\textrm{RPZ2}}$
do not give a significant improvement. 
The bound $B_{\textrm{RPZ3}}$ performs better than $B_{\textrm{MU}}$, 
but is typically worse than $B_{\textrm{Maj1}}$.
Note that some authors define the entropies with log base two
while in the present work we in general use the natural logarithm instead. 
In Table 1 we however switch to $\log_{2}$ while presenting numerical comparison of all bounds for the special case $U=O_3$.

\begin{table}
\begin{tabular}{|c|c|}
\hline 
The bound & Approximate value\tabularnewline
\hline 
\hline 
$B_{\textrm{D}}$ & 0.425\tabularnewline
\hline 
$B_{\textrm{MU}}$ & 0.585\tabularnewline
\hline 
$B_{\textrm{CP1}}$ & 0.623\tabularnewline
\hline 
$B_{\textrm{CP2}}$ & 0.641\tabularnewline
\hline
$B_{\textrm{RPZ1}}$ & 0.649\tabularnewline
\hline 
$B_{\textrm{RPZ2}}$ & 0.649\tabularnewline
\hline 
$B_{\textrm{RPZ3}}$ & 0.676\tabularnewline
\hline 
$B_{\textrm{Maj1}}$ & 0.669\tabularnewline
\hline 
$B_{\textrm{Maj2}}$ & \textbf{0.688}\tabularnewline
\hline 
\end{tabular}
\\
\caption{Comparison between numerical values of all bounds in the case of $U=O_3$ in the $\log_2$
 units used in \cite{Coles}.}
\end{table}

\begin{figure}
\includegraphics[scale=0.18]{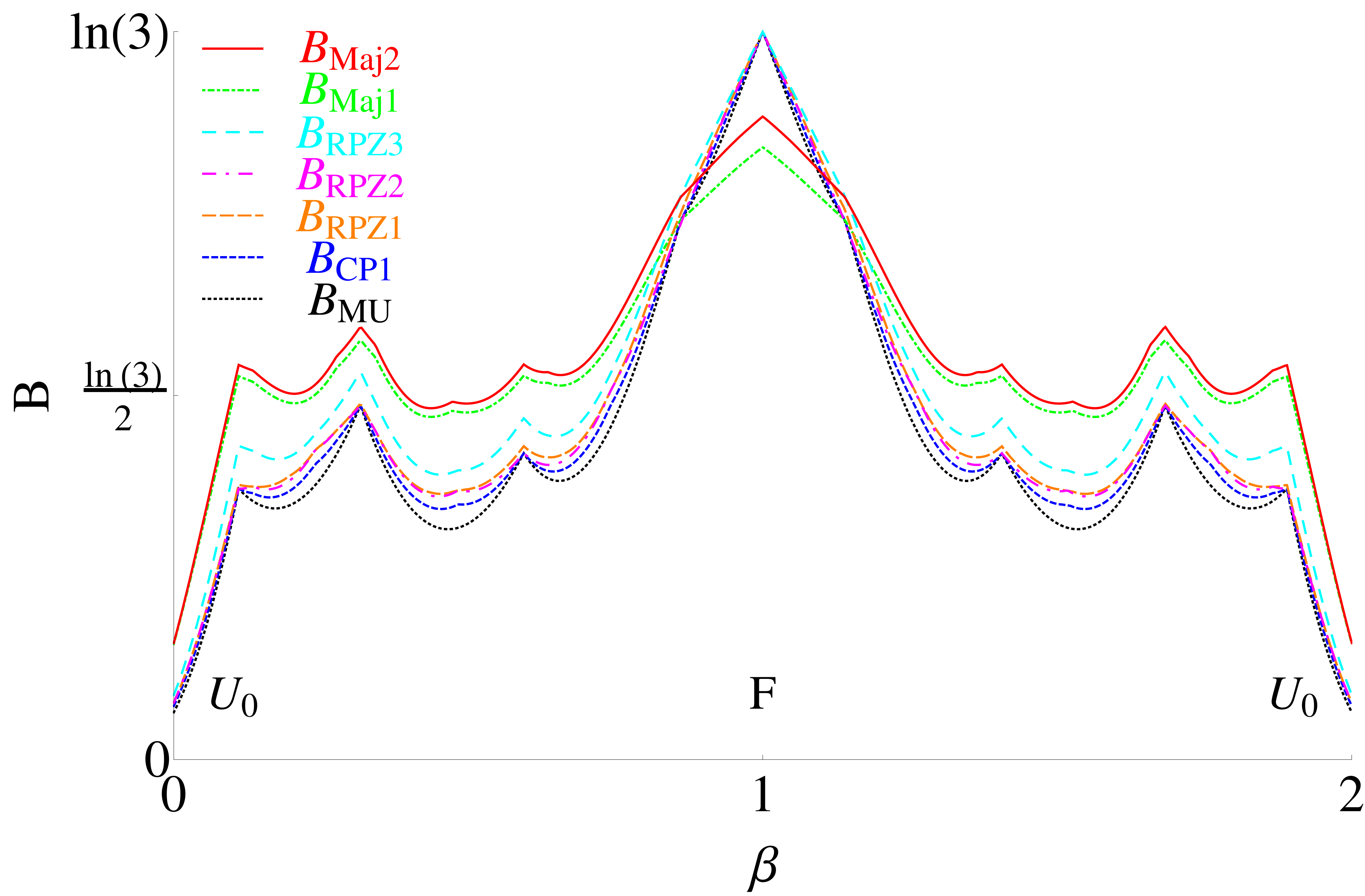}
\caption{(Color online) Bounds for a family of unitary matrices \eqref{eqn:family2}.}
\label{fig:fig2}
\end{figure}

In Fig.~\ref{fig:fig2} we plot the bounds for $3 \times 3$ matrices given by
\begin{equation} \label{eqn:family2}
U_\beta = (F_3)^\beta \exp( i (1-\beta) H ) .
\end{equation}
Here  $F_d$ denotes the Fourier matrix of order $d$,
so that $(F_d)_{jk}=\exp(2 \pi i jk/d) / \sqrt{d}$, while our model Hamiltonian 
$H$ reads
\begin{equation}
H = 
\left(
\begin{smallmatrix}
0 & 1 & 2 \\
1 & 0 & 2 \\
2 & 2 & 0
\end{smallmatrix}
\right).
\end{equation}
This family thus interpolates between the Fourier matrix $F_3$ and $U_0 = \exp(i H)$. 
In this case the direct-sum majorization bound is substantially better than the
Maassen-Uffink bound, while considering  matrices laying far away from the Fourier matrix.

\begin{figure}
\includegraphics[scale=0.18]{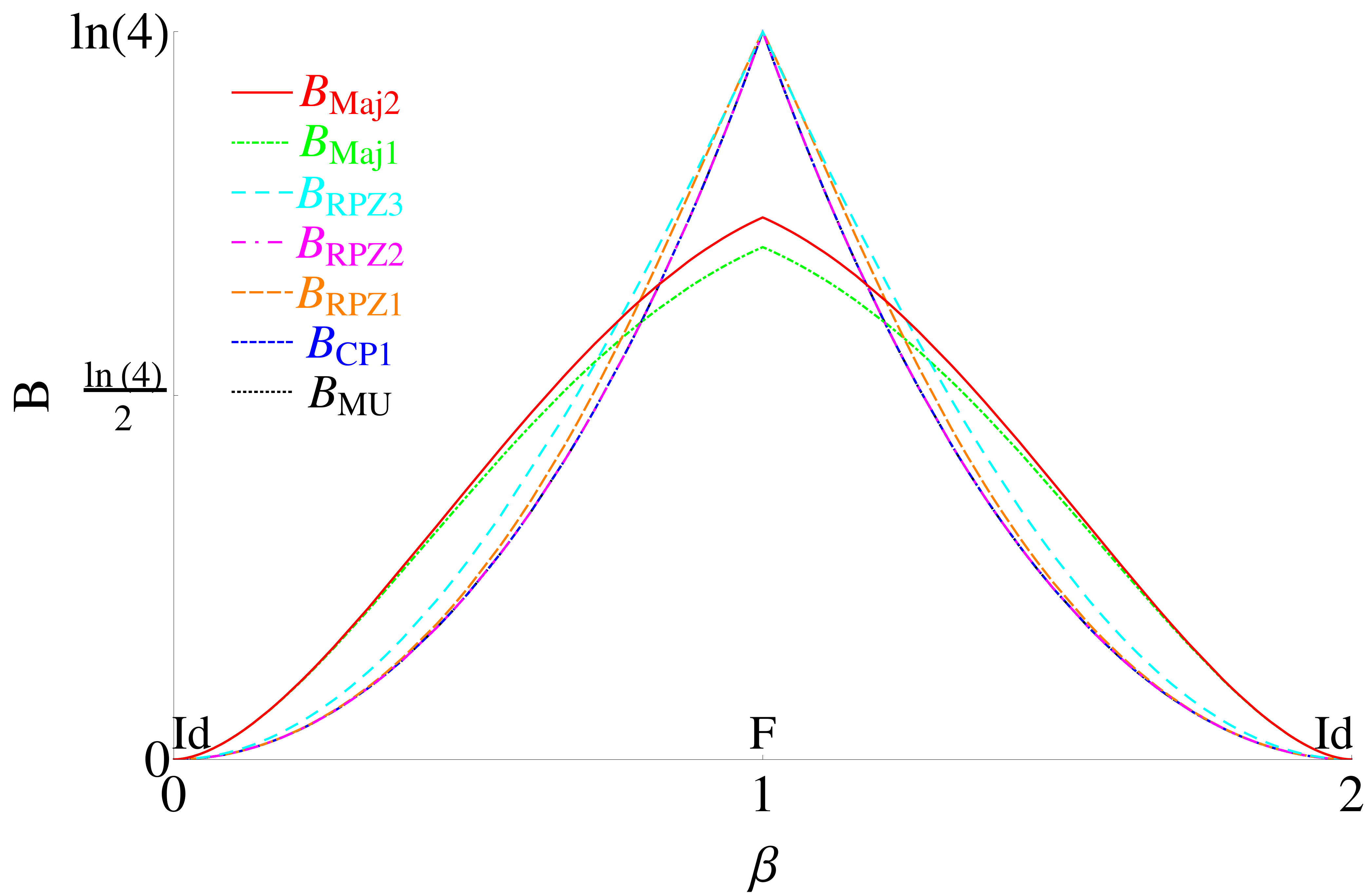}
\caption{(Color online) Bounds for a family of unitary matrices $(F_4)^\beta$
 interpolating between the identity and the Fourier matrix for $\beta \in [0,1]$.}
\label{fig:fig3}
\end{figure}

In Fig.~\ref{fig:fig3} we study the family of matrices which interpolates
between the identity and the  Fourier matrix $F_4$, namely $U(\beta) = (F_4)^\beta$.
Similarly to the prior case the direct-sum majorization relation provides  a better
bound for matrices which are distant from the Fourier matrix, while in its
neighborhood the $B_{\textrm{RPZ3}}$ bound gives the best estimate.


\section{Several measurements}

Majorization entropic uncertainty relations derived in Section \ref{sec_maj2}
can be easily generalized to the case of an arbitrary number
of $L$ measurements. The problem is now given by a collection of arbitrary $L$
unitary matrices, $U^{(1)}, \dots, U^{(L)}$, one of which is usually
set to the identity.

Let  $\{\ket{u^{(j)}_i}\}$ be $i$-th column of the matrix $U^{(j)}$.
We shall consider an entropic uncertainty relation of the form
\begin{equation}
\label{nowy0}
\sum_{i=1}^L H \bigr( p^{(i)}\bigl)
\ge \ B,
\end{equation}
where $p^{(j)}_i = |\scalar{u_{i}^{(j)}}{\psi}|^2$. In order to find a candidate for the bound $B$ we shall
 introduce a majorizing vector in a similar manner to the approach  presented
in Section \ref{sec_maj2}.

First of all,  we define new coefficients $\mathcal{S}_k$ being maximal squares of
norms calculated for the rectangular matrices of size $d \times ( k+1)$ formed by $k+1$ columns taken from the concatenation of all $L$ matrices $\{U^{(j)}\}_{j=1}^L$, i.e.
\begin{equation}
\mathcal{S}_k = \max \{
\sigma_1^2(\ket{u_{i_1}^{(j_1)}},\ket{u_{i_2}^{(j_2)}}, \dots,
\ket{u_{i_{k+1}}^{(j_{k+1})}})
\},
\end{equation}
where the maximum ranges over all subsets $\{
(i_1,j_1),(i_2,j_2),\dots, (i_{k+1},j_{k+1}) \}$
of cardinality $k+1$ of set $\{1,2,\dots,d\} \times \{1,2,\dots,L\}$.
It is easy to realize that $\mathcal{S}_0 = 1$, as all vectors
$u_i^{(j)}$ are normalized.

In the case $L=2$ one gets
\begin{equation}
\mathcal{S}_k = 1+ s_k,
\end{equation}
with $s_k$ defined in Eq. (\ref{sk}).

Using the methods presented in \cite{my} one can show that for any vector
$\ket{\psi} \in \mathbb{C}^d$,
\begin{equation}
\{p_i^{(j)}\}_{i,j=1}^{d,L} \prec \{1, \mathcal{S}_1-1, \mathcal{S}_2
- \mathcal{S}_1, \dots\}.
\end{equation}
The above observation leads to the following entropic uncertainty relation 
\begin{equation}
\label{nowy}
\sum_{i=1}^L H \bigr( p^{(i)}\bigl)
 \ge
-\sum_{i=1}^{d L} (\mathcal{S}_i - \mathcal{S}_{i-1})
\ln(\mathcal{S}_i - \mathcal{S}_{i-1})
\end{equation}

In the case of the R{\'e}nyi entropies with $\alpha<1$ and also the Tsallis entropies
one can easily formulate lower bounds similar to \eqref{nowy}.

\begin{figure}
\includegraphics[scale=0.18]{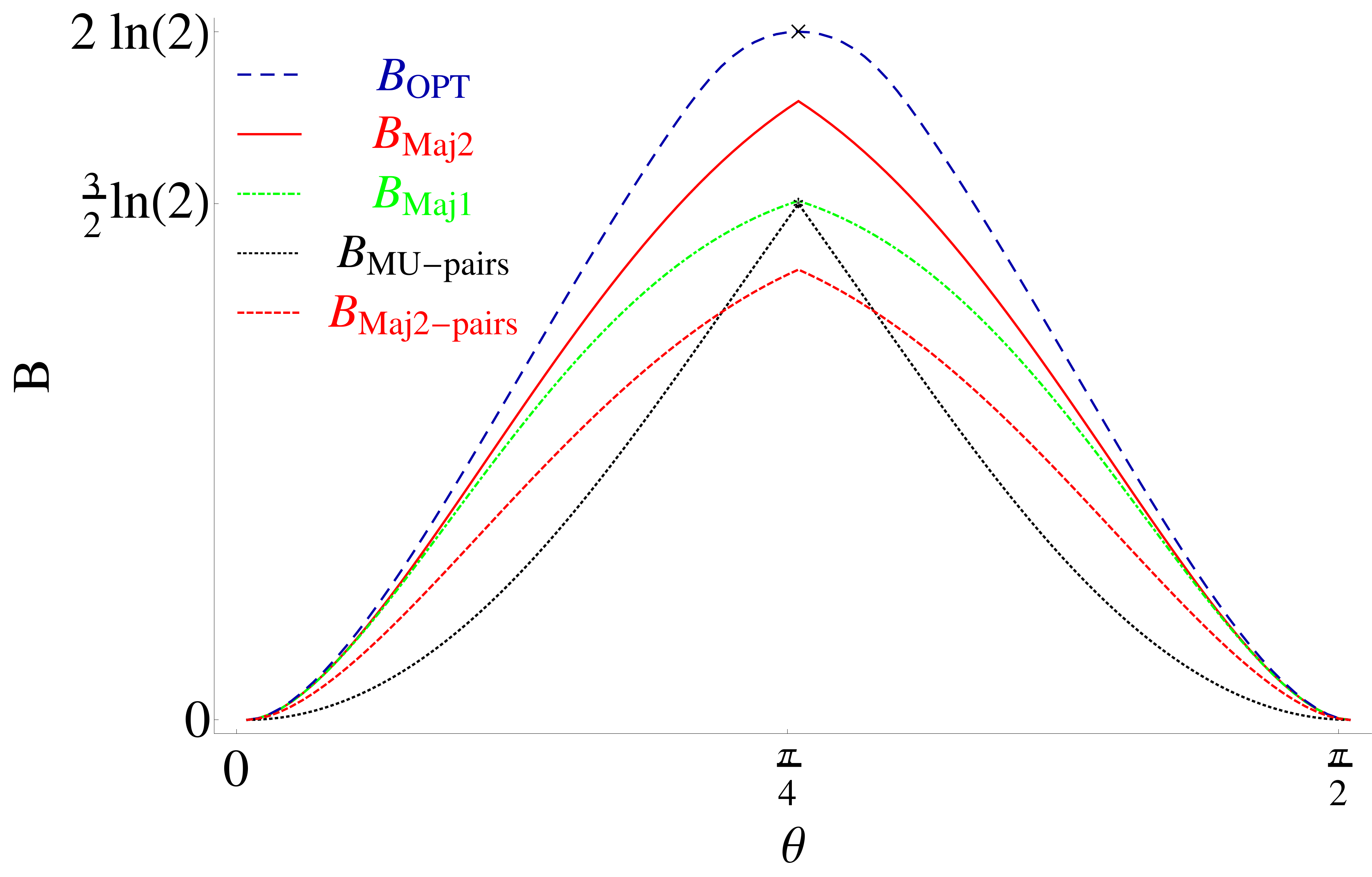} 
\caption{(Color online) Bounds for the sum of three entropies corresponding to  
    measurements in three bases in $\mathbbm{C}^2$ defined by \eqref{eqn:3-mubs}.
   Curve $B_{\rm OPT}$ represents the optimal bound obtained numerically.}
\label{fig:3-mubs}
\end{figure}

To illustrate, the case of more than two measurements we shall employ families which
interpolate between identical and mutually unbiased bases.
In Fig.~\ref{fig:3-mubs} we consider 3 bases, represented by the columns of three unitary matrices of order two
\begin{equation} \label{eqn:3-mubs}
U^{(1)}={\mathbbm{I}}_2, 
\ \
U^{(2)}=
\left(
\begin{smallmatrix}
\cos \theta & \sin \theta \\
\sin \theta & -\cos \theta
\end{smallmatrix}
\right)
,
\ \
U^{(3)}=
\left(
\begin{smallmatrix}
\cos \theta & \sin \theta \\
i \sin \theta & -i \cos \theta
\end{smallmatrix}
\right).
\end{equation}
In the case of $\theta=0$ we obtain bases which give the same measurements
probabilities, while in the case of $\theta = \pi /4$ the three bases are mutually unbiased.
In this case the sum \eqref{nowy0} of three entropies is larger than $2 \log 2$ -- the MUB bound of Sanchez \cite{Sanchez},
extended for mixed states in~\cite{CYGG11}. 
 In the neighborhood
of  $\theta = \pi /4$  the Maassen-Uffink bound calculated pairwise,
labeled by $B_{\rm MU-pairs}$, 
is  obviously larger than the direct-sum majorization bound calculated pairwise, $B_{\rm Maj2-pairs}$.
However, the direct-sum majorization bound \eqref{nowy}
performs better than the Maassen-Uffink bound calculated pairwise.

\begin{figure}
\includegraphics[scale=0.18]{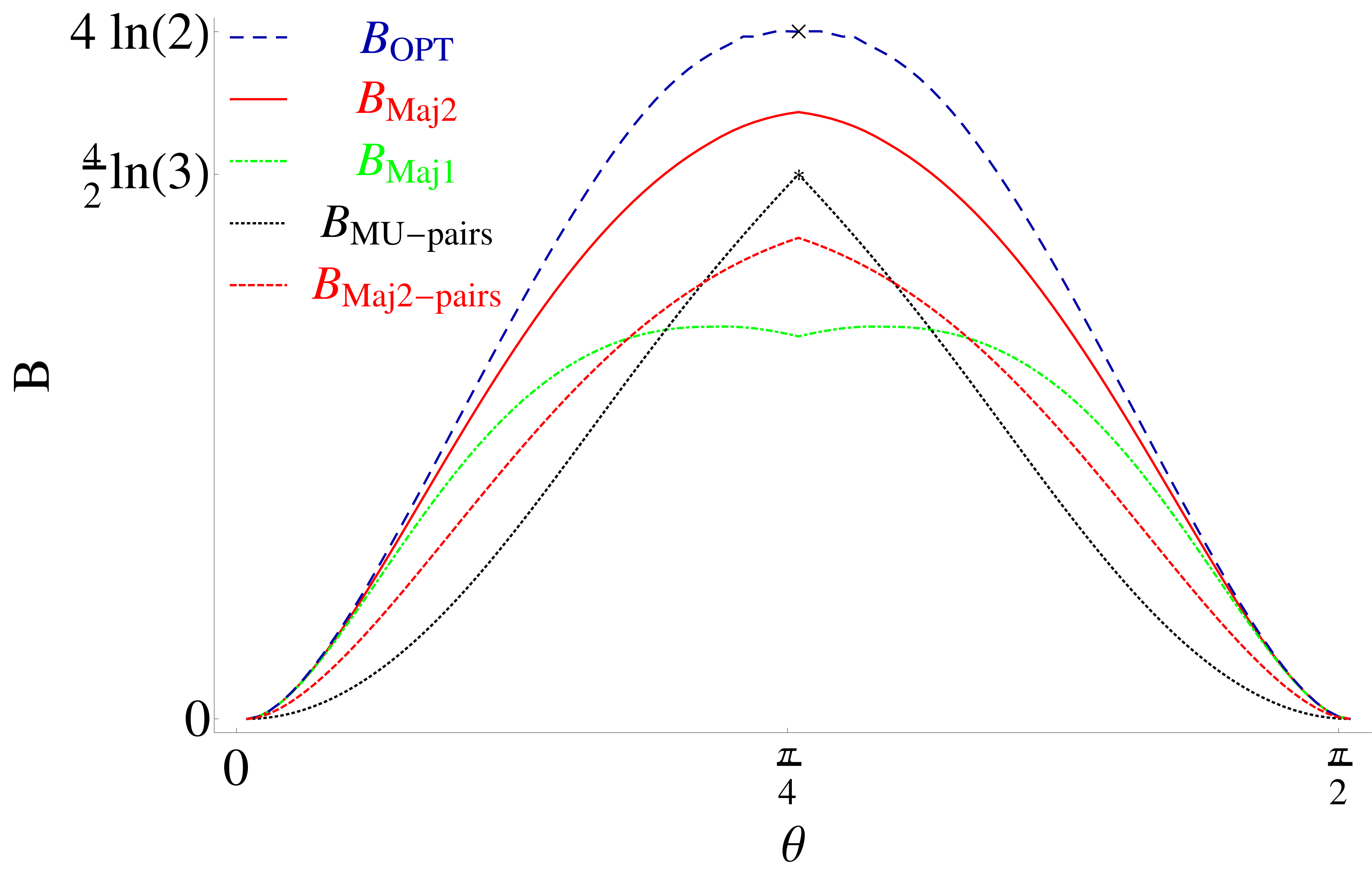} 
\caption{(Color online) Bounds for the sum of four entropies
related to four unitary matrices (\ref{eqn:4-mubs},\ref{eqn:4-mubs-cd}).
Sign ($\times $) represents the MUB result of Sanchez \cite{Sanchez}.
}
\label{fig:4-mubs}
\end{figure}

A  similar behavior is shown in Fig.~\ref{fig:4-mubs}
obtained for $L=4$ bases of size  $d=3$ defined by
\begin{equation}\label{eqn:4-mubs}
U^{(1)}={\mathbbm{I}}_3, 
\ \
%
U^{(2)}= (F_3)^{4 \theta/\pi} ,
\end{equation}
\begin{equation}\label{eqn:4-mubs-cd}
U^{(3)}= E (F_3)^{4 \theta/\pi},
\ \
U^{(4)}= E^2 (F_3)^{4 \theta/ \pi},
\end{equation}
where $E = {\rm  diag} (1,\exp(i 2 \pi /3),\exp(i 2 \pi /3))$.
Note that for $\theta =0$ all matrices become diagonal and  correspond to the same basis,
while for $\theta=\pi/4$ the bases are mutually unbiased (MUB). 
In this case the direct-sum majorization bound is close to the optimal bound
obtained numerically. 

Observe that the direct-sum majorization bound (\ref{nowy}) 
valid for the collection of $L$ unitary matrices $U^{(i)}$
by construction gives results which are generically better and always not worse 
than using the same bound pairwise for all $L(L-1)/2$ pairs of 
unitary matrices $V_{ij}=U^{(i)\dagger} U^{(j)}$. This statement follows 
from the fact that in the latter case one performs optimization over a smaller set.

\section{Concluding remarks}

In this work we derived several families of universal bounds for the sum of Shannon
entropies corresponding to orthogonal measurements of a given quantum state $\rho$ of size $d$
in arbitrary $L$ bases. In the simplest case of $L=2$ the problem is set by specifying 
a single unitary matrix $U$ of order $d$.

 If absolute value of the largest entry of the matrix $U$ 
is significantly smaller than $1$, what is the case e.g. when $U$ is close to the Fourier matrix, 
the most accurate results are obtained by the hybrid bounds $B_\textrm{RPZ1}$ and $B_\textrm{RPZ3}$. In the opposite case, when $U$ contains some entries 
of modulus close to unity, the direct-sum majorization bound  (\ref{Dir})
is generically better than all other bounds. Since when $c_1$ is close to $1/d$ the bounds $B_\textrm{RPZ1}$ and $B_\textrm{RPZ3}$ 
seem to be not worse than other known bounds,
and the bound (\ref{Dir}) is always not worse than the tensor-product majorization bound established in \cite{my, oni}
it is then fair to say that the collection of results derived in this work, 
provides the best set of bounds currently available.

As reported in \cite{my} for $d=3,4,5$ the bound (\ref{maj1}) 
applied to a random unitary matrix of order $d$ is 
generically stronger than the relation (\ref{Mu})  of Maassen--Uffink.
However, it can be shown that for large $d$ the situation changes 
and the following relation \cite{ALPZ14} holds asymptotically
$\langle B_{\textrm{Maj1}} \rangle_U < \langle B_{\textrm{MU}} \rangle_U < \langle B_{\textrm{Maj2}} \rangle_U$,
so the bound (\ref{Dir}) becomes the strongest one.
Here $\langle B \rangle_U$ denotes the bound $B$ averaged over the set of unitary matrices of order $d$
distributed with respect to the Haar measure.  

Observe that all three new inequalities (\ref{bound1}, \ref{bound2}, \ref{bound3}) 
work for an arbitrary mixed state $\rho$. Furthermore, 
the analyzed bound for the sum of two entropies characterizing both measurements
is enlarged by the von Neumann entropy of the measured state, $S(\rho)= -\textrm{Tr}\rho\ln\rho$,
which is equal to zero for any pure state.
Although our hybrid bounds were proved in this work for the monopartite case,
folowing Coles and Piani~\cite{Coles} one can also formulate them
in a more general, bi-partite setup.

An analogous result concerning the generalization of the Maassen-Uffink bound for the mixed states
appeared in \cite{Berta,CYGG11} and very recently in \cite{KLJR14}.
The first term on the right hand side of (\ref{bounds_mixed}),
due to the non-commutativity of both measurement bases,
is called \cite{KLJR14} the {\sl quantum uncertainty}, 
while the second term, related to the degree of mixing and
referred as the {\sl classical uncertainty}, vanishes for pure states.

The bounds obtained with the help of the direct-sum majorization
can be easily generalized to the case of an arbitrary number of measurements.
Analyzing exemplary families of three unitary matrices of size $d=2$ and 
four matrices of size $d=3$ which interpolate between $L$ identity matrices
and the set of mutually unbiased bases  we show that
the method proposed is applicable in practice for any collection of orthogonal measurements
and generically provides stronger results than these known previously. 

Strong majorization entropic uncertainty relations, established in this work,
can be used complementarily for various problems in the theory of quantum information.
Specific applications include separability conditions and characterization of multipartite entanglement \cite{GL04}, 
estimation of mutual information \cite{Coles}
in context of the Hall's information exclusion principle \cite{Ha95,G++13}, 
and for improved witnessing of quantum entanglement and measurements in presence of quantum memory \cite{Berta}.

\acknowledgments

It is a pleasure to thank  Gustavo Bosyk, Patrick Coles, Yuval Gefen, Kamil Korzekwa,
Marco Piani and Alexey Rastegin for stimulating discussions and useful correspondence.  
Financial support by NCN grants number DEC-2011/02/A/ST2/00305 (K\.{Z})
and DEC-2012/04/S/ST6/00400 (ZP), and grant number IP2011 046871 of
the Polish Ministry of Science and Higher Education (\L{}R) are gratefully
acknowledged.

\appendix\section{Derivation of direct-sum majorization relations}

We shall first prove the following lemma. 
\begin{lemma}
For probability vectors $p$ and $q$ defined in \eqref{eqn:def-p-q} the following majorization relation holds
\begin{equation}
\label{A11}
\left(p_{1},p_{2},\dots,p_{d},q_{1},q_{2},\dots,q_{d}\right)\prec\left(1,s_{1},s_{2}-s_{1},\dots s_{d}-s_{d-1}\right).
\end{equation}
\end{lemma}
Note, that the right hand side of the above relation concerns the vector $W$ present in Eq. (\ref{WW}).

\begin{proof}
Let us denote $z=p\oplus q$. We necessarily have $z_{i}\leq1$ for
all $i$, so that the first element of the vector majorizing $z$
must be equal to $1$. If we now consider the sum o two different
elements $z_{i}+z_{j}$ it can be the sum of two probabilities $p$,
the sum of two probabilities $q$ (in both cases the sum is bounded
by $1$), or the mixed sum $p_{i}+q_{j}$ bounded by $1+s_{1}$.
In a similar fashion we obtain more general inequalities
\begin{equation}
z_{i_{1}}+z_{i_{2}}+\dots+z_{i_{k}}\leq1+s_{k-1},
\end{equation}
proven in \cite{my} which gives the relation~(\ref{A11}). 
\end{proof}

\medskip 

Now we can prove Theorem~\ref{th:D-S-M}.
Let us first
consider the case $\alpha<1$. We shall begin with a simple observation,
that for $x\geq0$, the function $\ln(1+x)$ is subadditive, i.e.
\begin{equation}
\ln(1+x)+\ln(1+y)\geq\ln(1+x+y).
\end{equation}
Since $\alpha<1$ the sum $\sum_{i}p_{i}^{\alpha}$ as well as its
$q$-counterpart are greater than $1$. By putting $x=\sum_{i}p_{i}^{\alpha}-1$
and $y=\sum_{j}q_{j}^{\alpha}-1$ we find that 
\begin{equation}
H_{\alpha}(p)+H_{\alpha}(q)\geq\frac{1}{1-\alpha}\ln\left(\sum_{i}p_{i}^{\alpha}+\sum_{i}q_{i}^{\alpha}-1\right).
\end{equation}
Using the fact that the sum $\sum_i  x_i ^{\alpha}$ is Schur-concave for $\alpha < 1$
and utilizing the direct-sum majorization relation $z\prec\{1\}\oplus W$ we immediately
get the inequality (\ref{Dir}). The case of $\alpha=1$ is even simpler,
since we do not need to resort to subadditivity. Schur-concavity of
the Shannon entropy together with the fact that $-1\ln1=0$ gives
the desired result.

In order to prove (\ref{Renyi2}) we rewrite the sum of two R{\'e}nyi entropies as
\begin{equation}
H_{\alpha}(p)+H_{\alpha}(q)=H_{\alpha}(r)
\end{equation}
where $r = p \otimes q$. We then use the fact that the geometric mean is smaller than or equal to the arithmetic mean  
\begin{equation}
\sum_{i}p_{i}^{\alpha}\sum_{j}q_{j}^{\alpha}\leq\frac{1}{4}\left(\sum_{i}p_{i}^{\alpha}+\sum_{j}q_{j}^{\alpha}\right)^{2},
\end{equation}
and apply the direct-sum majorization relation. 

The proof of the Theorem~\ref{th:Tsallis}, concerning the Tsallis entropy, relies on the fact that
for $\alpha>0$, $\alpha\neq1$ and $x_{i}\geq0$ the function $\left(1-\alpha\right)^{-1}\sum_{i}x_{i}^{\alpha}$
is Schur concave. We have 
\begin{equation}
T_{\alpha}\left(p\right)+T_{\alpha}\left(q\right)=\frac{1}{1-\alpha}\left(\sum_{i}z_{i}^{\alpha}-2\right)\geq T_{\alpha}(W),
\end{equation}
where the last inequality follows from $z\prec\{1\}\oplus W$.

\section{Proof of $B_\textrm{Maj2}\geq B_\textrm{Maj1}$}

The vector $Q^{\left(d-1\right)}$ present in (\ref{maj1}) has the general form \cite{my}
\begin{equation}
Q^{\left(d-1\right)}=\left(R_{1},R_{2}-R_{1},\ldots,R_{d}-R_{d-1}\right),
\end{equation}
with 
\begin{equation}
R_{i}=\left(\frac{1+s_{i}}{2}\right)^{2},
\end{equation}
and $s_i$ defined in Eqn.~\eqref{sk}.
The above vector $Q^{\left(d-1\right)}$, as well as the vector $W$ are not necessarily sorted in a decreasing order,
that is why the proof of the majorization relation
$W = (s_1,s_2-s_2,s_3-s_2, \dots, s_d-s_{d-1})\prec Q^{\left(d-1\right)}$ is not straightforward.

Let us first show that for any unitary matrix $U$, we have $W_{1}\geq W_{k}$
for $k=2,\dots,d$, i.e. 
\begin{equation}
\label{B2}
s_{k}-s_{k-1}\leq s_{1}.
\end{equation}

By definition (\ref{sk}), there exist a matrix $A$ of dimension $n\times m$, being a submatrix
of $U$ and two normalized vectors
$|x\rangle$ and $|y\rangle$ of size $n$ and $m$  respectively, such that $n+m=k+1$ and 
\begin{equation}
s_{k}=|\bra{x}A\ket{y}|.
\end{equation}
Without loss of generality we can assume that $n\geq m$.
Since the vector $\ket{x}$ is normalized there exist $i\in\{1,2,\dots,n\}$,
such that $|x_{i}|\leq1/\sqrt{n}$ (by permuting the indices we may
assume that $|x_{1}|\leq1/\sqrt{n}$). Next we write 
\begin{equation}
\begin{split}
s_{k} & =|\bra{x}A\ket{y}|=\left|\sum_{i=1}^{n}\sum_{j=1}^{m}\overline{x_{i}}A_{ij}y_{j}\right|\\
&=
\left|\sum_{i=2}^{n}\sum_{j=1}^{m}\overline{x_{i}}A_{ij}y_{j}+\overline{x_{1}}\sum_{j=1}^{m}A_{1j}y_{j}\right|\\
&=\left|\bra{\tilde{x}} \tilde{A} \ket{y}+\overline{x_{1}}\scalar{a_1}{y}\right|,
\end{split}
\end{equation}
where $\bra{\tilde{x}}$ denotes the bra vector $\bra{x}$ without the first component,
 while  $\tilde{A}$ denotes the matrix $A$ without its first row
 and $\bra{a_1}$ denotes the first row of $A$. 
Next, we bound $s_k$ with the help of the triangle inequality, the Cauchy-Schwarz inequality, and the
fact that the overlap with normalized vectors does not exceed the largest 
singular value: 
\begin{equation}
\begin{split}s_{k} & =|\bra{\tilde{x}} \tilde{A}\ket{y}+\overline{x_{1}}\scalar{a_1}{y}|\leq|\bra{\tilde{x}} \tilde{A}\ket{y}|+|x_{1}||\scalar{a_1}{y}|\\
 & \leq\sqrt{1-|x_{1}|^{2}}\sigma_{1}(\tilde{A})+|x_{1}|\|a_1\|.
\end{split}
\end{equation}
Now using the fact, that $\max_{i,j}|U_{ij}|=s_{1}$ we get, that
$\|a_1\|\leq\sqrt{m}s_{1}$, we also have $x_{1}\leq\frac{1}{\sqrt{n}}$
and $\sqrt{1-|x_{1}|^{2}}\leq1$, and by definition $\sigma_{1}(\tilde{A})\leq s_{k-1}$.
We obtain 
\begin{equation}
\begin{split}s_{k} & \leq\sqrt{1-|x_{1}|^{2}}\sigma_{1}( \tilde{A})+|x_{1}|\|a_1\|\\
 & \leq\sigma_{1}(\tilde{A})+\frac{1}{\sqrt{n}}\|a_1\|
\leq s_{k-1}+\frac{\sqrt{m}}{\sqrt{n}}s_{1}
\\ & 
\leq s_{k-1}+s_{1},
\end{split}
\end{equation}
which directly implies Eq.(\ref{B2}). 

\medskip 

To prove the majorization relation  $W\prec Q^{\left(d-1\right)}$ 
we note that for $k\in\{2,\dots,N\}$ we have $Q^{\left(d-1\right)}_{k}\leq W_{k}$,
to see it we write 
\begin{equation}
\begin{split}
R_k - R_{k-1}
& =\frac{1}{4}\left(s_{k}^{2}-s_{k-1}^{2}+2(s_{k}-s_{k-1})\right)\\
 & =\frac{1}{4}(s_{k}-s_{k-1})\left(s_{k}+s_{k-1}+2\right)\\
 & \leq(s_{k}-s_{k-1}).
\end{split}
\end{equation}
The last inequality follows from the fact that $s_{k-1}\leq s_{k}\leq1$.

Using  inequality (\ref{B2}) we know, that $W_{1}\geq W_{k}$ and $Q^{\left(d-1\right)}_{1}\geq Q^{\left(d-1\right)}_{k}$
for any $k\in\{1,2,\dots,N\}$. This implies that the sums of the smallest
elements obey  the following inequalities inequalities:
\begin{eqnarray}
(W^{\uparrow})_{1} & \geq &(Q^{\uparrow})_{1} \nonumber \\
(W^{\uparrow})_{1}+(W^{\uparrow})_{2} & \geq &(Q^{\uparrow})_{1}+(Q^{\uparrow})_{2}\\
(W^{\uparrow})_{1}+(W^{\uparrow})_{2}+(W^{\uparrow})_{3} & \geq &(Q^{\uparrow})_{1}+(Q^{\uparrow})_{2}+(Q^{\uparrow})_{3} \nonumber \\
\vdots \nonumber
\end{eqnarray}
where $(W^{\uparrow}),(Q^{\uparrow})$ are vectors $W,Q^{\left(d-1\right)}$ ordered increasingly.
Since the total sum of both vectors is the same (and equal to $1$) we obtain 
the desired majorization relation $W\prec Q^{\left(d-1\right)}$.

\end{document}